\documentclass[aps,twocolumn,showpacs,floatfix,prl]{revtex4}
\usepackage{bbding}
\usepackage{epsfig}
\usepackage{graphicx}
\usepackage{dcolumn}
\usepackage{bm}
\usepackage{ulem}
\usepackage{lipsum}
\usepackage{amssymb}
\usepackage{amsmath}
\usepackage{dsfont}
\usepackage{soul,color}
\newcommand{\be}{\begin{equation}}
\newcommand{\ee}{\end{equation}}
\newcommand{\bea}{\begin{eqnarray}}
\newcommand{\eea}{\end{eqnarray}}

\newcommand{\sgn}{{\rm sign}}

\begin{document}
\title{Perfect spin filter by periodic drive of a ferromagnetic quantum barrier}

\author{Daniel Thuberg$^1$, Enrique Mu\~noz$^1$, Sebastian Eggert$^2$ and Sebasti\'an A. Reyes$^1$}

\affiliation{$^1$Instituto de F\'isica and Centro de Investigaci\'on en Nanotecnolog\'ia y Materiales Avanzados, Pontificia Universidad Cat\'olica de Chile, Casilla 306, Santiago 22, Chile}
\affiliation{$^2$Physics Department and Research Center OPTIMAS, University of Kaiserslautern, D-67663 Kaiserslautern, Germany}

\begin{abstract}
{ We consider the problem of particle tunneling through a periodically driven ferromagnetic 
quantum barrier connected to two leads.   The barrier is modeled
 by an impurity site representing a 
ferromagnetic layer or
quantum dot in
a tight-binding Hamiltonian with
a local magnetic field and an AC-driven potential, which is solved using the 
Floquet formalism.  The repulsive interactions in the quantum barrier are also taken into account.
Our results show that the time-periodic potential causes sharp resonances of perfect transmission
and reflection, which can be tuned by the frequency, the driving strength, and the 
magnetic field. 
We demonstrate that a device based on this configuration could act as a 
highly-tunable spin valve
for spintronic applications.}  
%even in the presence of a local Coulomb interaction. }
\end{abstract}

\maketitle

%{\it Introduction.}
The problem of a magnetic impurity embedded in a metallic matrix has been extensively studied over many years, particularly since the discovery of the Kondo effect \cite{Kondo_64}. The Kondo
or sd model that captured the low-energy physics in the local moment regime of such systems
was soon proved to be equivalent to the Anderson model \cite{Anderson_61}. In the early years, the main interest was focused on the thermodynamic properties of such systems, and analytical solutions  \cite{Yosida.70,Yamada.75,Yamada.76,Yamada.79,Zlatic.83} as well as numerical methods such as numerical renormalization group \cite{Costi.94c,Costi.10} have been extensively applied to deal with strong electronic correlations. More recently, the problem has recovered
interest for its non-equilibrium and transport properties\cite{Beenakker.91,iftikhar2015two}, particularly since the possibility to construct 
{ tunable impurities in }
experimental systems such as semiconductor quantum dots or single-molecule transistors became available  \cite{GoldhaberGordon.98,Scott.09,Scott.13,perrin2015single}. 

The vast majority of the theoretical work related to the non-equilibrium problem 
{ considers}
the steady-state conductance as a function of a constant bias voltage \cite{Schiller.95,Majumdar.98,Oguri.01,Oguri.05,Hewson.05,Sela.09,Pletyukhov.12} and its universal aspects \cite{Doyon.06}. A more difficult, however exciting problem, arises when the system is subjected to a {periodically driven potential}. 
%Another interesting experimental scenario that can be described by one-dimensional chains
%of the Hubbard type is provided by quasi-1D cuprates \cite{Schiappa_2012,Lee_013}. In such materials,
%experimental evidence via inelastic X-ray scattering shows that spin (spinons) and charge (holons)
%resolved are carried as independent quasi-particle excitations along the quasi-one dimensional lattice.

Besides
the solid state scenario, periodically driven tight-binding systems can be realized with
optical lattices and cold atom systems\cite{Bloch_05}. These types of systems have provided an exciting scenario,
not only as experimental simulations of solid-state models, but also to test configurations
with potential new features \cite{Bloch_05,Zhang_014,Greschner_015,aidelsburger2015measuring,douglas2015quantum,ponte2015many,ott14,shaking1,shaking2,agarwal17,PhysRevLett.118.260602}.

\begin{figure}[t]
% \left
\includegraphics[width=\columnwidth]{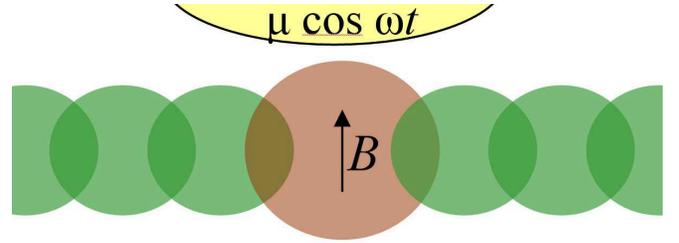}
 \caption{Schematic setup of a ferromagnetic quantum barrier %(layer or dot) 
subjected to a periodic drive and connected to two leads.}
\label{quantumdot}
\end{figure}

{ In this work, we study the transmission of electrons through a ferromagnetic quantum barrier
with a driven gate potential and 
connected to two leads as illustrated in Fig.~\ref{quantumdot}.   }
We consider electronic excitations of opposite spin traveling along a tight-binding chain with nearest neighbor hopping amplitude $J$. 
A periodically driven gate with angular frequency $\omega$ and amplitude $\mu$ is imposed 
upon the quantum barrier at the central site { together with a local
magnetic field $B$.} The possibility of a local electron-electron interaction of strength $U$ 
{and a local potential energy $\epsilon_d$ on the barrier}
are also included. The resulting Hamiltonian reads ($\hbar=1$)
\bea
\label{Eq:Hamiltonian}
H &=& -J \sum_{i,\sigma} (c_{i,\sigma}^\dagger c_{i+1,\sigma}{\phantom{\dagger}} + c_{i+1,\sigma}^\dagger c_{i,\sigma}{\phantom{\dagger}}) \\ 
&&+ \sum_{\sigma}\left(\epsilon_d -\mu \cos (\omega t) -\sigma b \right) c_{0,\sigma}^\dagger c_{0,\sigma}{\phantom{\dagger}} + U n_{0, \uparrow} n_{0, \downarrow}\nonumber
\eea
{in standard notation where $b=\frac{\mu_B}{2} B$.
It has been established that single
particle transmission through a driven impurity without magnetic field can give 
rise to sharp resonances, where the transmission vanishes completely even for 
infinitesimally small driving amplitudes \cite{reyes2017transport,thuberg2016quantum}.  We now show that the addition of a 
magnetic field and the local
potential energy not only make the transmission spin-dependent but 
also cause an entirely new 
effect of  {\it perfect transmission} for special parameters.
The combination of such ballistic transmission and total 
reflection for different spin-dependent parameters therefore 
opens the possibility to construct a perfect spin filter,} 
that may be very attractive for spintronics applications. 

To calculate the transmission probability for each spin channel we will 
find steady-state solutions to the Schr\"odinger equation
\be
(H(t) -i \partial_t) \left|\Psi(t)\right> = 0. \label{Eq:Schrodinger}
\ee
{using the Floquet formalism \cite{Hanggi_98} 
for a time-periodic Hamiltonian of the form 
$H(t) = H_0+2H_1\cos(\omega t)$ as in Eq.~(\ref{Eq:Hamiltonian}). 
The steady-state solution is a 
so-called Floquet state $\left|\Psi(t)\right>=e^{-i\epsilon t}\left|\Phi(t)\right>$,
which can be determined by the eigenvalue equation
\be
(H(t) -i \partial_t) \left|\Phi(t)\right> = \epsilon \left|\Phi(t)\right>. \label{Eq:eigenvalue}
\ee
where $\left|\Phi(t)\right> = \left|\Phi(t+2 \pi/\omega)\right>$ is time-periodic and $\epsilon$ is the quasienergy.}
Using the spectral decomposition 
\be
|\Phi(t)\rangle = \sum_{n=-\infty}^{\infty}e^{-i n \omega t} |\Phi_{n}\rangle, \label{Eq:spectral}
\ee
the eigenvalue equation  becomes
\be
H_0|\Phi_n\rangle + H_1(|\Phi_{n+1}\rangle+|\Phi_{n-1}\rangle) = (\epsilon +n\omega)|\Phi_{n}\rangle .
\ee
The repulsive Coulomb interaction $U$ on the barrier induces many-body correlations, 
which are technically difficult to deal with even in thermal equilibrium. 
As we show in the Appendix \ref{suppl} it is possible to develop a mean-field approach for Floquet systems
by using a time-dependent density on the barrier $\langle n_{0, \sigma}\rangle$
such that $n_{0, \uparrow} n_{0, \downarrow} \approx  \langle n_{0, \uparrow}\rangle n_{0, \downarrow} +  \langle n_{0, \downarrow}\rangle n_{0, \uparrow} -  \langle n_{0, \uparrow}\rangle \langle n_{0, \downarrow}\rangle$.
%where $\langle n_{0, \sigma}\rangle$ is the average occupation of spin $\sigma$ on site 0. 
%, that allows us to work in the single-particle picture, at the expense of introducing a non-linear, self-consistent condition.  When considering the effect of the Coulomb interaction, we shall use a mean-field approximation 
In this way it is possible to use a general single particle state with spin $\sigma$, which is defined by the coefficients for all modes
of the spectral decomposition
\be
|\Phi_{n}^\sigma\rangle = \sum_{j} \phi^\sigma_{j,n} c_{j,\sigma}^\dagger |0\rangle, \label{Eq:staten}
\ee
where $|0\rangle$ is the vacuum state. 
Inserting Eq.~(\ref{Eq:staten}) into the eigenvalue equation (\ref{Eq:eigenvalue}) results
in recursion relations for the amplitudes $\phi_{j,n}$. For the bulk ($j \neq 0$) we have
\be \label{Eq:bulk}
-J (\phi^\sigma_{j-1,n}+\phi^\sigma_{j+1,n}) = \bar{\epsilon}_{n, \sigma} \phi^\sigma_{j,n}, 
\ee
where 
\be
\bar{\epsilon}_{n, \sigma} = \epsilon + U \beta_{n, \sigma}  + n \omega .
\ee
Here we have defined {a mean-field parameter}
$\beta_{n, \sigma} = \sum_m \nu_{m, \sigma} \nu_{n-m, \bar{\sigma}}$
and
$\nu_{m,\sigma} = \sum_n \phi^{\sigma\ *}_{0,n} \phi^\sigma_{0,n+m}$, where we have used the 
notation $\bar{\sigma}=-\sigma$.  
{In contrast to ordinary mean field calculations, it is essential that the density on 
the driven quantum barrier is time-dependent (see Appendix \ref{suppl}).  This has the interesting 
consequence that all 
Floquet modes become coupled}  at the quantum barrier for $j=0$
\bea \label{Eq:amplitudes}
&-&J (\phi^\sigma_{-1,n}+\phi^\sigma_{1,n}) - \frac{\mu}{2}(\phi^\sigma_{0,n+1}+\phi^\sigma_{0
,n-1})
 \nonumber\\
&+& U \sum_m \nu_{m,\bar{\sigma}} \phi^\sigma_{0,n-m}
= (\bar{\epsilon}_n - \epsilon_d +\sigma b) \phi^\sigma_{0,n}.
\eea

%{\it Transmission coefficient.} We now proceed to calculate the exact transmission coefficient $T$ for two incoming particles of opposite spin with wavenumber $k_0$ for the mode $n=0$ corresponding to a quasi-energy $\bar{\epsilon} = -2J\cos k_0$ according to Eq.~(\ref{Eq:bulk}).
The time-periodic potential in the quantum barrier is not energy conserving and 
can cause scattering into other Floquet modes $n$.  For an incoming wave with 
 wavenumber $k_0$ for the mode
$n=0$ with quasi-energy $\bar{\epsilon} = -2J\cos k_0$,
the solution of 
Eq.~(\ref{Eq:bulk}) has the form
\bea 
|\Phi_n^\sigma\rangle &=& \sum_{j<0} \left[\delta_{n,0} A e^{i k_0 j} c_{j,\sigma}^\dagger + e^{-i k_n j} r_{n, \sigma} c_{j, \sigma}^\dagger \right] | 0 \rangle \nonumber \\  
&+& \sum_{j>0}e^{i k_n j} t_{n, \sigma}  c_{j, \sigma}^\dagger | 0 \rangle
+ E_{n,\sigma} c^{\dagger}_{0,\sigma} | 0 \rangle .
\eea
where the wavenumbers are given by $-2J\cos k_n = \bar{\epsilon} + n\omega$.
If $|\bar{\epsilon}+n\omega|<2J$, $k_n$ is real 
and the corresponding plane wave solutions are delocalized over 
the entire chain (unbound channels), which is always the case for the incoming wave $n=0$. 
For modes with $\bar{\epsilon}+ n\omega<-2J$, $k_n=i\kappa_n$ is imaginary and the solutions decay exponentially around the impurity (bound channels).
For $\bar{\epsilon}+ n\omega>2J$ the solutions decay and oscillate with a complex wavenumber $k_n=i\kappa_n+\pi$.
Using Eq.~(\ref{Eq:bulk}) it is easy to check that
\be
E_{n, \sigma}=t_{n, \sigma}=r_{n, \sigma}+\delta_{n,0} A, \label{rtn}
\ee
capturing the inversion symmetry of the lattice with respect to $j=0$.  
Inserting the amplitudes $\phi_{j,n}^{\sigma}$ that arise after Eq.~(\ref{rtn}) back into Eq.~(\ref{Eq:amplitudes}), we obtain a recursive relation for the coefficients $E_{n, \sigma}$:
\bea 
E_{n+1,\sigma} + E_{n-1,\sigma} &=& -  \frac{4 i\,J}{\mu} \sin k_n (E_{n,\sigma} - A \delta_{n,0})\nonumber\\
&& + \frac{2}{\mu} U \gamma_n
- \frac{2}{\mu}\sigma (b - {\sigma}\epsilon_d)E_{n,\sigma}, \label{Eq:recurrence}
\eea
which is the central equation that needs to be solved by requiring
convergence $E_{|n|\to \infty} \to 0$.  Here 
the influence of the interaction is captured by 
the term $\gamma_{n, \sigma} = \sum_m \nu_{m,\sigma} \phi^{\bar{\sigma}}_{0,n-m}$, 
which obviously depends on the total density of particles with opposite spin and can
be iteratively determined self-consistently.
%Note however that 
%particles of the same spin with different wavenumbers remain independent, so that an 
%arbitrary distribution of incoming amplitudes $A(k_0)$ can be considered.  
In the following, we assume 
an unpolarized incoming current 
composed of equal amplitudes for opposite spin.

For the transmission coefficient, it is useful to observe that 
that the current of the incoming wave (normalized to $|A|^2 \sin k_0$) has to equal the sum of all outgoing waves
\be 
\sum_{n} \left(|r_{n,\sigma}|^2 + |t_{n,\sigma}|^2 \right) \sin(k_n) = A^2 \sin(k_0).
\ee
Therefore, 
the total transmission can be expressed in terms of the solution for $E_{n,\sigma}$ 
\bea 
T_{\sigma} &=& \frac{1}{A^2}\sum_{n} T_{n,\sigma} = \frac{1}{A^2} \sum_{n} |E_{n,\sigma}|^2 \cdot \frac{\sin(k_{n})}{\sin(k_0)} = \frac{{\rm Re} E_{0,\sigma}}{A}\nonumber\\
&=& {\rm Re}\left[\frac{u_k}{u_k - \frac{i \mu}{2} \left(\frac{E_{1,\sigma}}{E_{0,\sigma}} + \frac{E_{-1,\sigma}}{E_{0,\sigma}} \right) -  i\, \sigma \tilde{b}_{\sigma} }\right], \label{transmission}
\eea
where we have used Eq.~(\ref{Eq:recurrence}) for $n = 0$ in the last line, with
$u_k=2J\sin k_0$ as the incoming particle velocity and 
\be
\tilde{b}_{\sigma} = b - \sigma(\epsilon_d + U\gamma_{0,\sigma}/E_{0,\sigma}).
\label{eq_btilde}
\ee

%we consider 
%\be 
%E_{0,\sigma} = \frac{A}{1 - \frac{i\mu}{2 u_k} \left(\frac{E_{1,\sigma}}{E_{0,\sigma}} + \frac{E_{-1,\sigma}}{E_{0,\sigma}} + \frac{2}{\mu} \sigma\tilde{b}_{\sigma} \right)},
%\ee
%with
%Particle number conservation yields 
%Thus, to calculate $T_{\sigma}$ we need to determine
%iteratively the occupation amplitude on the dot $E_{n,\sigma}$ considering the effect of the interaction ($\gamma_{n,\sigma}, \beta_{n,\sigma}$). 

\begin{figure}[t]
% \left
\includegraphics[scale=0.45]{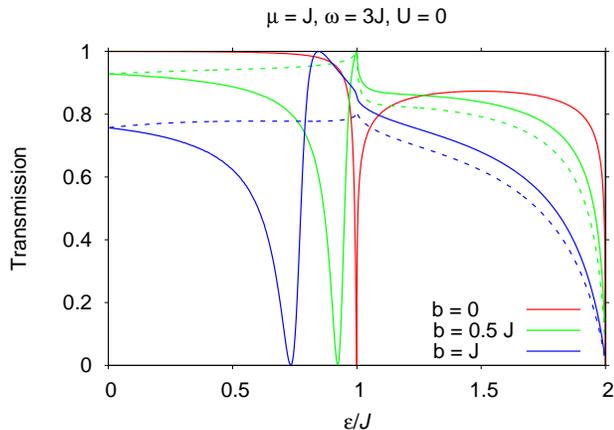}
 \caption{Transmission of both spin channels (solid line: spin up; dashed line: spin down) for a perturbation with $\omega=3J$, $\mu=J$ and various magnetic field strengths $b$ as function of $\epsilon$. The interaction is turned off ($U=0$) and $\epsilon_d=0$}.
\label{m_1_w_3_magn_div}
\end{figure}

Let us first consider the effect of a magnetic field $b$ without interactions $U=0$ and for
vanishing on-site energy $\epsilon_d=0$, as shown in Fig.~\ref{m_1_w_3_magn_div} as a function
of incoming energy $-2J < \epsilon < 2J$.  
Since the results are the same for $\epsilon \to -\epsilon$ and $\tilde{b}_\sigma \to -\tilde{b}_\sigma$, only the energy range
of the upper half of the band is shown.
Maybe the
most striking features are the points of complete reflection $T=0$ at 
certain energies $\epsilon$.
For $b=0$ these reflection points were linked to the phenomena of Fano resonances and are known to occur even 
for arbitrary small driving amplitudes $\mu \to 0$ at incoming energies   
$\epsilon \to  \omega -2J$ \cite{thuberg2016quantum,reyes2017transport}.   For $\sigma b< 0$ there are no such resonances, but 
for $\sigma b> 0$ the points of perfect reflection now shift
to lower energies and we observe a new
feature of {\it perfect transmission} at nearby incoming energies, which opens the possibilities to construct a 
perfect spin-filter as outlined below.
To estimate the locations of the zero transmission resonances, it is instructive to consider 
the recurrence relation in Eq.~(\ref{Eq:recurrence}) for $n\neq 0$, which can be written as  
$E_{n+1,\sigma} + E_{n-1,\sigma} = \alpha_n E_{n,\sigma}$ where 
\be
\alpha_{n} = \frac{2}{\mu} \left[-\sgn{(n)}\sqrt{(\epsilon+n\omega)^2-4J^2} - \sigma\tilde{b}_{\sigma}\right]. \label{alpha}
\ee
For small driving amplitudes $\mu \to 0$ the $\alpha_n$ grow beyond bounds, but a resonance 
condition $E_{0,\sigma}/E_{-1,\sigma} \to 0$ in Eq.~(\ref{transmission}) is still possible for $\alpha_{-1} \to 0$, so that 
the points of zero transmission are given by
\be
\epsilon \stackrel{\mu\to 0}{\longrightarrow} \omega - \sqrt{4J^2+\tilde{b}_{\sigma}^2}
\label{eq_Fano}
\ee
for $\sigma b > 0$. 
As mentioned above there are corresponding resonances for $\epsilon\to -\epsilon$ and reversed spin (or field).
However,  if the frequency is too small or the field is too large so that the expression in Eq.~(\ref{eq_Fano}) becomes negative, the resonances
are pushed outside the band and there will be no points of zero transmission for any energy.  
On the other hand, for any given incoming energy $\epsilon$ in the band it is possible to find a sufficiently high frequency so that Eq.~(\ref{eq_Fano}) can be
fulfilled.
%(SOME INTUITION BEHIND THE FORMULA)

%(HERE MORE ABOUT SPIN FILTER ETC)

\begin{figure}[t]
 \centering
\includegraphics[scale=0.45]{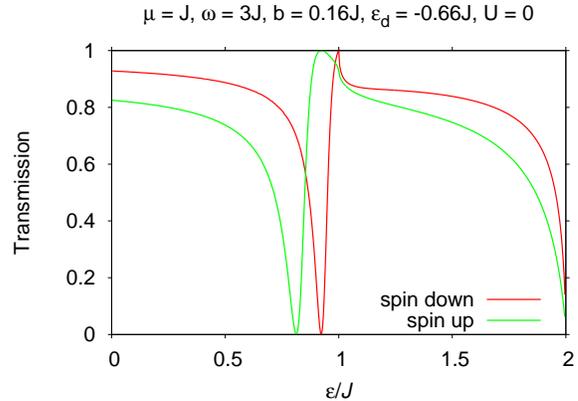}
 \caption{Transmission behavior of both spin channels for a perturbation with $\hbar \omega=3J$, $\mu=J$, a magnetic field strength $b=0.16J$ and a constant potential $\epsilon_d=-0.66J$ showing the possibility for a perfect spin filter.}
\label{fig_perfect_spin_filter}
\end{figure} 

The points of perfect transmission $T=1$ in Fig.~\ref{m_1_w_3_magn_div} are also linked to the Fano resonance, which has been studied for 
 static side coupled systems \cite{miroshnichenko2005engineering}.  In our case the effective magnetic field  $\sigma\tilde{b}_{\sigma}$ in 
Eq.(\ref{eq_btilde}) leads to a finite Fano asymmetry parameter and therefore a nearby point of perfect transmission.
Basically the reduced Zeeman energy enhances the local occupation at the impurity site and increases
transmission through the barrier.  This effect can also be achieved by a local potential energy $\epsilon_d$ as an additional tuning parameter.
In particular, it is straight-forward to choose a negative value of $\epsilon_d$ and a small positive value of $b$, so that the 
effective on-site energy $\sigma \tilde{b}_\sigma$ is attractive for both spin-channels but resulting  in a spin-dependent shift
in Eq.~(\ref{eq_Fano}).  In Fig.~\ref{fig_perfect_spin_filter} the parameters $b=0.16J$ and $\epsilon_d=-0.66J$ were chosen so that 
the transmission maximum for spin-up occurs at the same energy as the resonance of perfect reflection for spin-down.
This demonstrates that it is possible to create a perfect spin-filter by a combination of a static magnetic field 
and a local time-periodic potential.

\begin{figure}[t]
% \left
\includegraphics[scale=0.45]{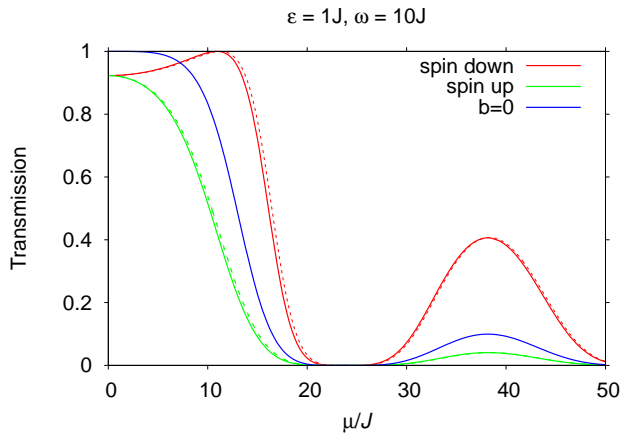}
 \caption{Transmission of both spin channels for a perturbation with $\omega=10J$ and a magnetic field of $b=0.5$ as function of $\mu$. The interaction is turned off ($U=0$) and $\epsilon_d=0$. The dashed lines correspond to the approximation from Eq.~(\ref{high_freq_approx_magn})}
\label{w_05_mu_3_b_025_U0}
\end{figure}

In the high frequency regime $\omega \gg J, \tilde{b}_{\sigma}$ the coefficients $\alpha_n$ in Eq.~(\ref{alpha}) can be expanded to first order in $\omega^{-1}$. The resulting (approximate) recurrence relation has an exact solution in terms of Bessel functions of the first kind $\mathcal{J}(x)$ \cite{thuberg2016quantum,della2007visualization}. Thus, in this regime we can obtain an analytical approximation for the transmission
% \be
% \frac{E_{-1,\sigma}}{E_{0,\sigma}} + \frac{E_{1,\sigma}}{E_{0,\sigma}} \approx \frac{J_{1 - \epsilon/\omega+b\sigma/\omega}(\mu/\omega)}{J_{-\epsilon/\omega+b\sigma/\omega}(\mu/\omega)} - \frac{J_{1 + \epsilon/\omega-b\sigma/\omega}(\mu/\omega)}{J_{\epsilon/\omega-b\sigma/\omega}(\mu/\omega)},
% \ee
% which leads to an approximate transmission of
\be
T_{\sigma} \approx \frac{u_k^2}{u_k^2+ \left(\frac{\mu}{2}\chi(\mu/\omega)+ \sigma\tilde{b}_{\sigma}\right)^2} \label{high_freq_approx_magn},
\ee
where $\chi(\mu/\omega)=\frac{\mathcal{J}_{1 - \epsilon/\omega-\sigma\tilde{b}_{\sigma}/\omega}(\mu/\omega)}{\mathcal{J}_{-\epsilon/\omega-\sigma\tilde{b}_{\sigma}/\omega}(\mu/\omega)} - \frac{\mathcal{J}_{1 + \epsilon/\omega+
\sigma\tilde{b}_{\sigma}/\omega}(\mu/\omega)}{\mathcal{J}_{\epsilon/\omega+ \sigma\tilde{b}_{\sigma}/\omega}(\mu/\omega)}$.
A comparison between this approximation and the exact result is shown in Fig.~(\ref{w_05_mu_3_b_025_U0}). Moreover, a first-order expansion of $\chi(\mu/\omega)$ in $\epsilon + \sigma\tilde{b}_{\sigma}$ leads to an analytical estimate for the location
of the point of perfect transmission $T_{\sigma} = 1$ at $\mathcal{J}_{0}(\mu/\omega)^2 = \sigma\tilde{b}_{\sigma}/\epsilon + 1$.
We see that tuning of the resonance locations is therefore also equally possible by changing $\mu$ or $\omega$ and
both calculations coincide almost perfectly already for $\omega = 10J$. 

Last but not least the effect of interactions $U \neq 0$ must be considered.  As can be seen from Eq.~(\ref{eq_btilde}) the main effect comes from
the renormalization of the effective magnetic field $\tilde{b}_{\sigma}$
by a spin-dependent contribution to the local potential energy $\epsilon_d + U\gamma_{0,\sigma}/E_{0,\sigma}$ in the mean field approximation.
Therefore, a similar effect as seen in Fig.~\ref{fig_perfect_spin_filter} is observed in this case, where two independent
Fano resonances appear for each spin channel. The position of the resonances is approximately given by Eq.~(\ref{eq_Fano}), with a small shift in comparison with the non-interacting case, which is self-consistently calculated
within the mean-field approximation through the parameter $U\gamma_{0,\sigma}/E_{0,\sigma}$. We notice that at high frequency $\omega$ 
and small amplitudes $\mu$, the occupation of the Floquet coefficients $n\neq 0$ becomes negligible, 
and hence in this limit one has $U\gamma_{0,\sigma}/E_{0,\sigma} \sim U \langle n_{0,\sigma}\rangle$, such that the effective site energy is approximately $\epsilon_d + U\langle n_{0,\sigma}\rangle$, which is precisely the Hartree-Fock expression
for a static impurity subject to a local interaction \cite{Hewson.93,Hewson}. This is consistent with the interpretation following Eq.~(\ref{high_freq_approx_magn}) that in the very high frequency regime, the oscillations average-out and 
the system can be effectively described by {\it static} fields and local potentials\cite{reyes2017transport,della2007visualization}.  
In particular, using an effective field on the barrier
 $b_{\rm eff} =  \tilde{b}_{\sigma} + \sigma \frac{\mu}{2}\chi(\mu/\omega)$ gives the same high frequency transmission coefficient
$T= b_{\rm eff}^2/(u_k^2 + b_{\rm eff}^2)$, i.e. the local potential must simply be shifted by $\frac{\mu}{2}\chi$.

In conclusion, we have analyzed the effect of a local time-periodic potential on the 
transport through a ferromagnetic quantum barrier including 
local potentials, fields, and interactions.
In contrast to static gating and filtering mechanisms, 
the periodic drive allows points of perfect transmission 
and complete reflection, which is useful for the generation of tunable spin-currents.
To achieve complete reflection for spin-down and perfect transmission for spin-up
as shown in Fig.~\ref{fig_perfect_spin_filter},
 two parameters must be tuned, such as the local static potential and the
driving frequency. 
 The technological implementation of such a perfect
spin filter requires relatively high frequencies
of the order of the incoming energy, which is normally 
the Fermi-energy relative to the band edge.  
Therefore, systems with low filling, small hopping, or very heavy effective
mass are most promising in this respect.   The geometry of the setup is of course not limited to a quasi one-dimensional array with a central quantum dot shown 
in Fig.~\ref{quantumdot}, since the effect can also be derived in the same way for any system where the 
transport channels go through a ferromagnetic layer with a tunable time-periodic potential.

%NOTE: In Fig. 4 use U=0.5. Then start with b=0, then b=0.1 and b=1

% \begin{figure}[h]
% % \left
% \includegraphics[scale=0.4]{w_3_mu_3_b_025_U0}
%  \caption{Transmission of both spin channels for a perturbation with $\mu=3J$ and $\omega=3$ and a magnetic field of $b=0.25$ as function of $\epsilon$. The interaction is turned off ($U=0$)}
% \label{w_3_mu_3_b_025_U0}
% \end{figure}

\begin{acknowledgments}
D. T. acknowledges financial support from CONICYT Grant No.63140250,  E. M. acknowledges financial support from Fondecyt (Chile) 1141146.  S.E. acknowledges support from the Deutsche Forschungsgemeinschaft (DFG) via the collaborative research centers SFB/TR173 and SFB/TR185.
\end{acknowledgments}
%\bibliographystyle{mystyle}
%\bibliography{AC_Driven}
\bibliographystyle{apsrev4-1}

\section{Appendix} \label{suppl}

Here we discuss in detail the mathematical formulation and intermediate steps leading to
the mean-field theory approximation and corresponding Floquet equations presented in the main body of the article.

For a system that is periodically driven, periodicity in time allows to express the solutions to the dynamical problem
in terms of a set of eigenfunctions $|\Psi_{}(t)\rangle = e^{-i\hbar^{-1}\epsilon_{}t}|\Phi_{}(t)\rangle$, with
$\epsilon_{}$ a Floquet eigenvalue, and an associated periodic eigenfunction $|\Phi_{}(t + T)\rangle = |\Phi_{}(t)\rangle$.
If one restricts the non-equivalent values of $\epsilon_{}$ to a first Brillouin zone $\epsilon_{} \in \left[-\pi\hbar/T, \pi\hbar/T \right]$,
then the periodic eigenfunction can be expanded as
\begin{eqnarray}
|\Phi_{}(t)\rangle = \sum_{n\in Z}e^{-i n \omega t} |\Phi_{n}^{}\rangle,
\label{eq1}
\end{eqnarray} 
where each stationary Floquet mode $|\Phi_{n}^{}\rangle$ is associated to an eigenvalue $\epsilon_{n} = \epsilon_{} + n\hbar\omega$
outside the first Brillouin zone.

Let us consider now the Hamiltonian described in the main body of the article,
\begin{eqnarray}
\hat{H} &=& -J\sum_{j,\sigma}\left( \hat{c}_{j+1,\sigma}^{\dagger}\hat{c}_{j,\sigma} + h.c.\right) + U\hat{n}_{0,\uparrow}\hat{n}_{0,\downarrow}\nonumber\\
&&
+ \left(\epsilon_d - \sigma b - \mu\cos(\omega t) \right)\sum_{\sigma}\hat{n}_{0,\sigma}
\label{eq2}
\end{eqnarray}
The exact treatment of the interaction would require a two-particle eigenbasis. Here, in order to obtain a simpler
physical interpretation of the transport properties, we decide to remain in the single-particle eigenbasis 
$|j,\sigma\rangle = \hat{c}^{\dagger}_{j\sigma}|0\rangle$, for $\{\hat{c}_{j\sigma},\hat{c}_{j'\sigma'}^{\dagger} \} = \delta_{\sigma\sigma'}\delta_{j,j'}$
Fermonic operators. Therefore, each stationary Floquet component in the periodic function defined by Eq.(\ref{eq1}) is expressed by a linear combination of the form
\begin{eqnarray}
|\Phi_{n}^{\sigma}\rangle = \sum_{j} \phi_{j,n}^{\sigma}|j,\sigma\rangle 
\label{eq_basis}
\end{eqnarray}
Therefore, we treat the Coulomb interaction in a mean-field theory (MFT) approximation, using the standard decoupling of the number operators
as follows
\begin{eqnarray}
U \hat{n}_{0,\uparrow}\hat{n}_{0,\downarrow} &\sim&  U \langle \hat{n}_{0,\uparrow}\rangle_{(t)}\hat{n}_{0,\downarrow} + U \langle \hat{n}_{0,\downarrow}\rangle_{(t)}\hat{n}_{0,\uparrow}\nonumber\\
&&- U   \langle \hat{n}_{0,\uparrow}\rangle_{(t)} \langle \hat{n}_{0,\downarrow}\rangle_{(t)}
\label{eq3}
\end{eqnarray}
In Eq.(\ref{eq3}), we have introduced the definition of the time-dependent expectation value of the number operators in the Floquet eigenstate $|\Phi_{}(t)\rangle$
\begin{eqnarray}
\langle \hat{n}_{0\sigma} \rangle_{(t)} &=& \langle \Phi_{}(t)|\hat{n}_{0,\sigma}|\Phi_{}(t)\rangle\nonumber\\
&=& \sum_{n_1,n_2\in Z} e^{-i(n_1 - n_2)\omega t}\langle \Phi_{n_2}^{}|
\hat{n}_{0,\sigma}|\Phi_{n_1}^{}\rangle
\label{eq4}
\end{eqnarray}
Notice that Eq.(\ref{eq4}) shows that the interaction couples different Floquet modes $|\Phi_{n}^{}\rangle$ through the dynamical expectation value
of the local number operators.
Let us now calculate the matrix elements involved, using the single-particle representation of the Floquet basis Eq.(\ref{eq_basis})
\begin{eqnarray}
\langle \Phi_{n_2}^{}|
\hat{n}_{0,\sigma}|\Phi_{n_1}^{}\rangle &=&
 \sum_{j_1,j_2,\sigma_1,\sigma_2} \left(\phi_{j_2,n_2}^{\sigma_2}\right)^{*}\phi_{j_1,n_1}^{\sigma_1}\nonumber\\
&&\times \langle j_2,\sigma_2|\hat{n}_{0,\sigma}|j_1,\sigma_1\rangle\nonumber\\
&=& \left(\phi_{0,n_2}^{\sigma}\right)^{*}\phi_{0,n_1}^{\sigma}
\label{eq5}
\end{eqnarray}
where we used the identity $\langle j_2,\sigma_2|\hat{c}_{0,\sigma}^{\dagger}\hat{c}_{0\sigma}|j_1,\sigma_1\rangle = \delta_{j_1,0}\delta_{j_2,0}\delta_{\sigma_1,\sigma}\delta_{\sigma_2,\sigma}$. Substituting Eq.(\ref{eq5}) into
Eq.(\ref{eq4}), reduces to the simpler expression
\begin{eqnarray}
\langle \hat{n}_{0\sigma} \rangle_{(t)} = \sum_{n\in Z} e^{-i n \omega t} \nu_{0,n}^{\sigma}
\label{eq6}
\end{eqnarray}
where we have defined the parameters
\begin{eqnarray}
\nu_{0,n}^{\sigma} = \sum_{m\in Z} \left(\phi_{0,m}^{\sigma}\right)^{*}\phi_{0,n+m}^{\sigma}
\label{eq_nu}
\end{eqnarray}
Using Eq.(\ref{eq6}), we can express the product of occupation numbers that appears in Eq.(\ref{eq3}) as
\begin{eqnarray}
\langle \hat{n}_{0,\uparrow}\rangle_{(t)}\langle \hat{n}_{0,\downarrow}\rangle_{(t)} &=& \sum_{n_1,n_2 \in Z} e^{-i(n_1 + n_2)\omega}
\nu_{0,n_1}^{\uparrow}\nu_{0,n_2}^{\downarrow}\nonumber\\
&=&\sum_{n\in Z} e^{-i n \omega t} \beta_{n}^{}
\label{eq7} 
\end{eqnarray}
Here, we have defined
\begin{eqnarray}
\beta_{n}^{} = \sum_{m\in Z}\nu_{0,n}^{\uparrow}\nu_{0,n-m}^{\downarrow}
\label{eq_beta}
\end{eqnarray}
Inserting the MFT terms into the Hamiltonian Eq.(\ref{eq2}), we obtain the effective single-particle MFT Hamiltonian
\begin{eqnarray}
&&\hat{H}_{MFT}(t) =  -J\sum_{j,\sigma}\left( \hat{c}_{j+1,\sigma}^{\dagger}\hat{c}_{j,\sigma} + h.c.\right) \nonumber\\
&& + \sum_{\sigma}\left[\left(\epsilon_d -\sigma b  
- \mu\right)\cos\omega t  + U\sum_{n\in Z} e^{-i n \omega t}
\nu_{0,n}^{\bar{\sigma}}\right]
\hat{n}_{0,\sigma}\nonumber \\
& & -\sum_{n\in Z} e^{-i n \omega t}\beta_{n}^{}(t)
\nonumber\\
\label{eq_HMFT}
\end{eqnarray}
The eigenvalue equation for this MFT effective Hamiltonian is
\begin{eqnarray}
\hat{H}_{MFT}(t)|\Phi_{}(t)\rangle = \left(\epsilon_{} + n\hbar\omega\right)|\Phi_{}(t)\rangle
\label{eq_eigenMFT}
\end{eqnarray}
Projecting this equation onto a single-particle state of the basis $\langle i,\sigma'|$, we have 
\begin{eqnarray}
& & \sum_{n\in Z} e^{-i n \omega t} \sum_{j,\sigma}\phi_{j,n}^{\sigma} \\&&\times\left( \langle i,\sigma'|\hat{H}_{MF}(t)|j,\sigma\rangle
- (\epsilon_{} + n\hbar\omega)\delta_{ij}\delta_{\sigma',\sigma}\right) = 0\nonumber
\label{eq8}
\end{eqnarray}
From the orthogonality of the set $\left\{ e^{-i n \omega t} \right\}_{n\in Z}$, we finally obtain the set of finite-differences equations
\begin{eqnarray}
&&-J\left( \phi_{i+1,n}^{\sigma} + \phi_{i-1,n}^{\sigma} \right) - \left(\epsilon_{} + n\hbar\omega + \beta_{n}^{} \right)\phi_{i,n}^{\sigma}\nonumber\\
&-&\delta_{i,0}\left[\left(\epsilon_d + \sigma b \right)\phi_{0,n}^{\sigma} + \frac{\mu}{2}\left(\phi_{0,n+1}^{\sigma} + \phi_{0,n-1}^{\sigma} \right) \right]\nonumber\\
&+&\delta_{i,0}U\sum_{m\in Z}\nu_{0,m}^{\bar{\sigma}}\phi_{0,n-m}^{\sigma} = 0
\label{eq9}
\end{eqnarray}
whose numerical and analytical solution is developed and discussed for different physical
scenarios in the main body of the article.

\end{document}